\documentclass[11pt]{article}
\usepackage{moriond,epsfig}

\def\Journal#1#2#3#4{{#1} {\bf #2}, #3 (#4)}

\def\NIMA{{\em Nucl. Instrum. Methods} A}

\def\PLB{{\em Phys. Lett.}  B}
\def\PRL{\em Phys. Rev. Lett.}

\def\APP{\em Astropart. Phys.}
\def\SP{\em Sol. Phys.}
\def\ApJ{\em ApJ}

\def\be{\begin{equation}}
\def\ee{\end{equation}}
\def\bea{\begin{eqnarray}}
\def\eea{\end{eqnarray}}

\begin{document}
\vspace*{4cm}
\title{EVIDENCE FOR TIME-VARYING NUCLEAR DECAY DATES: EXPERIMENTAL RESULTS AND THEIR IMPLICATIONS FOR NEW PHYSICS}
\author{E. FISCHBACH}
\address{Department of Physics, Purdue University, 525 Northwestern Ave,\\West Lafayette, Indiana, 47907 USA\\ephraim@purdue.edu}

\author{ J.H. JENKINS }
\address{School of Nuclear Engineering, Purdue University, West Lafayette, Indiana 47907 USA}

\author{P.A. STURROCK}
\address{Center for Space Science and Astrophysics, Stanford University, Stanford, California 94305  USA}

\maketitle\abstracts{
Unexplained annual variations in nuclear decay rates have been reported in recent years by a 
number of groups.  We show that data from these experiments exhibit not only variations in 
time related to Earth-Sun distance, but also periodicities attributable to solar 
rotation.  Additionally, anomalous decay rates coincident in time with a series of solar 
flares in December 2006 also point to a solar influence on nuclear decay rates.  This 
influence could arise from some flavor of solar neutrinos, or through some other objects 
we call ``neutrellos'' which behave in some ways like neutrinos.  The indication that 
neutrinos or neutrellos must interact weakly in the Sun implies that we may be able to 
use data on time-varying nuclear decay rates to probe the interior of the Sun, a 
technique which we may call``helioradiology''.}

\section{Introduction}\label{Intro}
The widely held view that nuclear decay rates, along with nuclear masses, are fundamental 
constants of nature has been challenged recently by reports from various groups of 
periodic variations in nuclear decay rates.\cite{jen09a,jen09b,fis09,stu10a,jav10,stu10b,stu11a,stu11b}  Following the discovery of natural 
radioactivity by Becquerel in 1896 an intense effort was mounted to determine whether 
nuclear decay rates were in fact constant.  By 1930 Rutherford, Chadwick, and Ellis 
concluded that``the rate of transformation of an element has been found to be a 
constant under all conditions.\cite{rut30} Subsequent work by many groups has supported 
this conclusion, except for decays proceeding by electron capture.  In such decays the 
decay rate depends on the overlap of the electron wavefunction and the nucleus, and 
this can be slightly modified by subjecting the decaying nucleus to extreme 
pressure, or by modifying its chemical environment.\cite{eme72,hop74,hah76,dos77,nor01,oht04,lim06}

Notwithstanding the impressive body of evidence supporting the conventional view that the decay 
rate $\lambda = ln2/T_{1/2}$ of an unstable isotope is an intrinsic property of that isotope, there are growing indications of small ($\mathcal{O}\left( 10^{-3}\right)$) time-dependent variations in the decay rates of some nuclei.  In Sec.~\ref{sec:exper} we summarize the existing experimental evidence for these variations, along with the arguments against some claims that the variations are simply the results of local environmental influences on the detector systems.  What emerges from these considerations is a picture in which the observed time-dependent variations are in fact occurring in the decay process itself, in response to time-dependent solar perturbations.  Although a detailed mechanism is not yet available to explain how the Sun influences radioactive decays, previous work\cite{fis09} along with the discussion in Sec.~\ref{sec:neut} below, provide a general framework for a future theory.

\section{Experimental Results}\label{sec:exper}

In this section we summarize the experimental evidence for a solar influence on nuclear decay processes.  A more detailed discussion of the various experiments can be found in the accompanying paper by Jenkins \textit{et al}.\cite{jen11}
The interest of our group in time-dependent nuclear decay rates began with our attempt to understand an annually-varying periodic signal in the decay rate of $^{32}$Si reported by Alburger, et al.\cite{alb86} in a 4-year experiment at the Brookhaven National Laboratory (BNL).  A subsequent examination of the literature revealed a $\sim$15-year experiment at the Physikalisch-Technische Bundesanstalt (PTB) in Germany in which data from $^{226}$Ra exhibited a similar annual variation.\cite{sie98}  Coincidentally, these two experiments overlapped in time for approximately two years, and the data from these experiments during this period were quite similar in both amplitude and phase.\cite{jen09b,fis09}  Further exploration of the literature has uncovered other experiments in which periodic effects in various decays were reported.  These include the results of Falkenberg\cite{fal01}, Parkhomov\cite{par05,par10}, Baurov et al.\cite{bau07}, Ellis\cite{ell90}, and Shnoll et al.\cite{shn98a,shn98b}.

The observation of periodic effects in what had been previously been thought of as random
decay data motivated a series of experiments by our group at Purdue, chiefly focused on the electron
capture process: $\rm{e}^- + ^{54}\rm{Mn} \rightarrow \nu_e + ^{54}\rm{Cr} + \gamma\left(834.8~\rm{keV}\right)$.
Our apparatus was operating during a series of solar flares in November and December of 2006. On 13 December 2006
a major solar flare erupted at 02:37 UT (21:37 EST 12 December) which coincided with a $\sim7\sigma$ drop in the measured $^{54}$Mn
counting rate.\cite{jen09a} A smaller flare with a large coronal mass ejection on 17 December 2006 also coincided with a decrease in the $^{54}$Mn counting rate. Subsequently, an examination of data acquired during December 2008
revealed a correlation between a change in the measured $^{54}$Mn count rate and a solar storm on the far side of the Sun.

The correlations between observed changes in measured $^{54}$Mn count rates and solar flares are significant
for several reasons: 
\begin{enumerate}
\item They now reinforce the inference, drawn from the annual periodicities in the BNL and PTB data, that these periodicities arise from the annual variation of the Earth-Sun distance $R$ due to the ellipticity of the Earth's orbit. 
\item Since the flares erupt and subside over fairly short time-scales (typically minutes to hours), any apparent correlation between decay data and solar activity cannot plausibly be attributed to environmental effects on the detector systems in question due to a local change in temperature, pressure, humidity, etc.\cite{jen10}
\item Finally, in all the cases we have observed, there is a precursor signal in which the $^{54}$Mn count rate begins to change $\sim$1 day before the solar event. This observation raises the possibility of establishing an ``early-warning'' system for potentially dangerous impending solar storms, whose damaging effects on astronauts; communications, navigation, defense and other satellites; and power grids and other electronic infrastructure could thus be prevented.\cite{jen09a}
\end{enumerate}

The observation in decay data of time-dependent influences attributable to the Sun (either from a change in $1/R^2$ or via a solar flare), raises the question of whether other time-dependent signals could be present in decay data associated, for instance, with solar rotation. This could happen if the sources of whatever influences were affecting decay rates were not distributed homogeneously throughout the Sun, for which there were earlier indications.\cite{stu08} Further analysis of the BNL, PTB, and Parkhomov data sets has indeed revealed evidence of a $\sim$32 d periodicity, which can be interpreted as evidence for an East-West asymmetry in the Sun, perhaps associated with a slowly rotating solar core.\cite{stu09} Additionally, the BNL and PTB data sets also revealed evidence of a $\sim$173 d periodicity, similar to the Rieger periodicity, which arises from retrograde waves in a rotating fluid.\cite{stu11a} Finally, recent work by our group has shown that the apparent phase-shift of the maximum count rate noted in BNL, PTB, and other data sets from what would be expected from the variation in $1/R^2$ alone could be attributed to a North-South asymmetry in the Sun.\cite{stu11b} While this observation does not directly deal with solar rotation, it does support the assumption that there are asymmetries in the Sun whose presence can be detected via periodicities in decay data. In this way decay data may allow us to probe the interior of the Sun via a new technique which we may refer to as ``helioradiology''.

\section{Towards a Mechanism: Neutrinos and Neutrellos}\label{sec:neut}

Although the evidence for a solar influence on nuclear decay rates is quite compelling, 
what is lacking is a mechanism through which this influence can be transmitted.   Elsewhere\cite{fis09}
we explore in detail a mechanism based on an interaction between solar neutrinos 
and decaying nuclei.  Here we broaden that discussion to address the question of whether 
nuclear decay rates are affected by the Sun through some generic particles which we 
call ``neutrellos'', which may or may not be the same as neutrinos.  In what follows we 
describe some of the properties that we would like neutrellos to possess in order to 
account for existing data.

\begin{enumerate}
\item{}The solar flare of 13 December 2006 at 02:37UT was coincident in time with a local minimum (dip) in the $^{54}$Mn counting rate.\cite{jen09a}  Since this dip occurred at 21:37 EST in our laboratory, this suggests that neutrellos must be capable of passing unimpeded through the Earth at essentially the speed of light.
\item{}Although the solar flare was of short duration and occurred without warning, the decay rate of $^{54}$Mn began to decrease much earlier, approximately 40 hours before the flare.  This ``precursor signal'' suggests that neutrellos originated from a region below the surface of the Sun and reached us before the actual flare because the Sun is effectively transparent to neutrellos, but not to photons.
\item{}As the Sun rotated, the region on the surface of the Sun from which the 13 December 2006 flare originated, region 930, dropped over the West Limb of the Sun on 17 December and hence was no longer visible via X-rays.  Nevertheless, a significant drop in the $^{54}$Mn count rate was detected on 22 December, suggesting that neutrellos were reaching the Earth from the far side of the Sun by passing through the Sun.  This again implies that the Sun is transparent to neutrellos, at least to some degree.
\item{}Our $^{54}$Mn experiment also detected a solar event on 16 December 2008 which coincided with a storm on the far side of the Sun.  This reinforces the assumption that the Sun is relatively transparent to neutrellos.  
\item{}The phase of the annual variation in decay rates seen in a number of experiments is shifted from what would be expected from the annual variation of $1/R^{2}$, where $R$ is the Earth-Sun distance.\cite{jen09b,fis09}  However, we have shown recently\cite{stu11b} that if there is a North-South asymmetry in the emission of neutrellos from the Sun, then the resulting contribution to the annual phase could explain the observed data.  Interestingly, evidence for a North-South asymmetry was observed in data from the Homestake solar neutrino experiment.\cite{stu98}
\item{}As noted previously, there is evidence that nuclear decay data are modulated not only by the variation of $1/R^{2}$ and the presence of a North-South asymmetry in the Sun, but also by the rotational motion of the Sun.  Evidence for rotational modulation is based on the presence in the decay data of periodicities of $\sim32$ d and $\sim173$ d, the latter being analogous to the well-known Rieger periodicity.  Since similar evidence for rotational modulation has been noted previously in data from the Homestake and GALLEX neutrino experiments\cite{stu08}, this suggests that our hypothetical neutrellos may actually be neutrinos.
\item{}The $\sim{}32$ d rotational modulation mentioned above points to a source where the rotation rate is slower than that of the radiative and convection zones in the Sun.  The fact that we observe a solar influence from a source below the radiative zone indicates that the putative neutrellos experience only slight (or no) scattering or absorption in travelling the outer layers of the Sun.
\item{}The very existence of a signal in nuclear decay data for rotational modulation by the Sun implies that neutrello production by the Sun is anisotropic.  If neutrellos were in fact neutrinos, then the rotational modulation could arise from the resonant spin flavor precession (RSFP)\cite{akh91} effect induced by a strong magnetic field deep in the solar interior.  This mechanism, which assumes that neutrinos have a non-zero transition magnetic moment, is supported by existing neutrino data from Super-Kamiokande.\cite{stu08}
\end{enumerate}
 
Although the preceding considerations are compatible with the inference that neutrellos are in fact neutrinos, there is at least one major difference:  to account quantitatively for existing experimental data the interaction strength of neutrellos with decaying nuclei must be significantly greater than the strength of the known interactions of neutrinos with protons, neutrons, electrons, or with other neutrinos as described by the standard electroweak model.  As an example, to produce a fractional peak-to-trough variation in tritium of order 10$^{-3}$ (which is the nominal value suggested by the BNL, PTB, and Falkenberg data) requires an input of energy $\Delta{}E\approx{}5eV$.  Although this is small on the scale of the  $\mathcal{O}(1 MeV)$ energies carried by incoming solar neutrinos, a value of $\Delta{}E$ this large is more characteristic of an electromagnetic interaction than a weak interaction.  This can be seen in another way by picturing solar neutrinos or neutrellos affecting nuclear decays by transferring a momentum $\Delta{}p\approx{}\Delta{}E/c$ via a scattering process with an effective cross section $\sigma$,

\begin{equation}
\sigma \equiv \frac{1}{\phi}\frac{\left(\Delta{}N/N\right)}{\Delta{}t}.
\label{eq:sigma}
\end{equation}

\noindent{}Here $\phi$ is the presumed flux (or change in flux) of solar neutrinos or neutrellos responsible for inducing a fractional change $\Delta{}N/N$ in the number of decays over a time interval $\Delta{}t$.  Evidently the smallest estimate of $\sigma$ will result from the largest assumed value for $\phi$ for which we adopt the known solar flux $\phi=6\times 10^{10}cm^{-2}s^{-1}$.  Using the flare data of Jenkins and Fischbach\cite{jen09a} we estimate $\left(\Delta{}N/N\right)/\Delta{}t \approx 2.6\times 10^{-11}s^{-1}$ per atom, and hence $\sigma \approx 4.3\times10^{-22} cm^2$.
By way of comparison, the Thomson cross section for photon scattering of electrons is
$\sigma_T = \left(8\pi{}r_o^2/3\right)=6.6\times10^{-25} cm^2$, where $ r_o=2.82\times10^{-13}cm$ is the classical electron radius.

The implication of the above calculations, that neutrinos could influence decaying nuclei through an interaction of electromagnetic strength, is likely incompatible with existing data on $\nu_s-e$, $\nu_s-p$, and $\nu_s-n$ interactions ($\nu_s$ = solar neutrino), but could be compatible with a possible $\nu_s$ -- $\nu_e$ interaction coupling a generic solar neutrino to an emitted $\nu_e$ from beta decay or electron capture. On the other hand,
a much broader range of possibilities is available for nutrello couplings, and these may be accessible experimentally through appropriate ``fifth force'' experiments.\cite{fis99} Additional constraints on a possible influence of $\bar{\nu}_e$ on radioactive decay follow from an elegant reactor experiment by de Meijer \textit{et al}.\cite{dem11}

\section{Discussion and Outlook}\label{sec:disc}

From the discussion in the previous sections several conclusions emerge: There is by now overwhelming evidence of anomalous and unexpected time-dependent features present in the count rates of various nuclei. Although some early criticism claimed that these features were merely experimental artifacts arising form the response of the detection systems to local seasonal changes in temperature, pressure, humidity, etc.,\cite{sem09} it now appears that the observed effects are intrinsic to the decay process itself. This follows from the detailed analysis of Jenkins, Mundy and Fischbach\cite{jen10} of the detector systems used in the BNL, PTB and Purdue experiments, and also from observation in multiple data sets of time-dependent features for which there is no known ``environmental'' cause.\cite{jen11}

Although the preceding discussion, along with the analysis in Section \ref{sec:exper}, suggests that the decay process is being influenced in some way by the Sun, there is at yet no detailed mechanism to explain how this influence comes about. Our discussion of neutrinos and neutrellos is an attempt to frame a future theory by outlining some of the specific characteristics that it should possess, given the limited experimental data currently available.

Evidently, more experimental data on a variety of different isotopes are needed before we can realistically expect to understand how the Sun influences radioactive decays. To start with, it is clear that there should be no expectation that time-dependent effects will show up in all decays, or that they should be detected at the same level when they are present. This follows by noting that the same details of nuclear structure that are responsible for the wide range of half-lives, from fractions of a second to tens of billions of years, will likely produce a range of responses to any solar influence. Moreover, if the Sun is in fact the source of the time-dependent effects observed in nuclear decays, its influence cannot be assumed to be constant in time. The well known $\sim11$ year solar cycle is but one example of a time-dependent solar feature whose affects would not be constant, or even periodic, over the duration of a typical laboratory experiment. For this reason, experiments on the same isotopes carried out at different times may not exhibit the same features. It is thus likely that ``helioradiology'' will be an important tool in studying the Sun, while at the same time creating new methods for studying neutrino (or neutrello) physics.

\section*{Acknowledgments}
The work of PAS was supported in part by the NSF through Grant AST-06072572, and that of EF was supported in part by U.S. DOE contract No. DE-AC02-76ER071428.


\end{document}